\documentclass{article}

\usepackage{amssymb}
\usepackage{amsmath}
\usepackage{amsbsy}
\usepackage{graphicx}
\usepackage{slashed}
\usepackage{ulem}
\usepackage{color}
\usepackage{psfrag}
\usepackage[nospace]{cite}

\newcommand{\bea}{\begin{eqnarray}}
\newcommand{\eea}{\end{eqnarray}}
\newcommand{\e}{{\rm e}}

\begin{document}

\begin{flushright}
MPP-2010-31\\[7mm]
\end{flushright}

\begin{center}
{\bf \Large Decay of a Yukawa fermion at finite temperature and
applications to leptogenesis}\\[4mm]


Clemens P.~Kie\ss ig$^\star$
\footnote{E-mail: \texttt{ckiessig@mpp.mpg.de}} ,
Michael Pl\"umacher$^\star$
\footnote{E-mail: \texttt{pluemi@mpp.mpg.de}}, 
Markus H.~Thoma$^\dagger$
\footnote{E-mail: \texttt{mthoma@mpe.mpg.de}}, 

\vspace*{0.5cm}
$^\star$ \it
Max-Planck-Institut f\"{u}r Physik (Werner-Heisenberg-Institut),\\
F\"ohringer Ring 6, D-80805 M\"unchen, Germany\\[0.1cm]
$^\dagger$ \it Max-Planck-Institut f\"ur extraterrestrische Physik,\\
Giessenbachstra\ss e, D-85748 Garching, Germany

\vspace*{0.4cm}
\end{center}

\begin{abstract}
We calculate the decay rate of a Yukawa fermion in a thermal bath
using finite temperature cutting rules and effective Green's functions
according to the hard thermal loop resummation technique. We apply
this result to the decay of a heavy Majorana neutrino in
leptogenesis. Compared to the usual approach where thermal masses are
inserted into the kinematics of final states, we find that deviations
arise through two different leptonic dispersion relations. The decay
rate differs from the usual approach by more than one order of
magnitude in the temperature range which is interesting for the weak
washout regime. We discuss how to arrive at consistent finite
temperature treatments of leptogenesis.
\end{abstract}

\section{Introduction}

Leptogenesis \cite{Sakharov:1967dj,Fukugita:1986hr} is an extremely
successful theory in explaining the baryon asymmetry of the universe
by adding three heavy right-handed neutrinos $N_i$ to the standard
model,
\begin{equation}
\delta {\mathcal L} = i \bar{N}_i \partial_\mu \gamma^\mu N_i - 
\lambda_{\nu,i
  \alpha} \bar{N}_i \phi^\dagger \ell_\alpha - \frac{1}{2} M_i
\bar{N}_i N_i^c + h.c.\: .
\end{equation}
with masses $M_i$ at the scale of grand unified theories (GUTs) and
Yukawa couplings $\lambda_{\nu,i \alpha}$ similar to the other
fermions. This also solves the problem of the light neutrino masses
via the see-saw mechanism without fine-tuning 
\cite{Minkowski,Yanagida:1979as,GellMann:1980vs}.

The heavy neutrinos decay into lepton and Higgs boson after inflation, the
decay is out of equilibrium since there are no gauge couplings to the
standard model. If the CP asymmetry in the Yukawa couplings is large enough, a
lepton asymmetry is created by the decays which is then partially
converted into a baryon asymmetry by sphaleron processes. As
temperatures are high, interaction rates and the
CP asymmetry need to be calculated using thermal field theory
\cite{Covi:1997dr,Giudice:2003jh,Besak:2010fb} rather than vacuum quantum 
field theory. However, in the conventional approach
\cite{Giudice:2003jh}, thermal masses have been put in by hand without 
investigating the validity of this approach in detail. We have
adressed this issue in \cite{Kiessig:2009cm} and found that
corrections arise through the occurence of two lepton
dispersion relations in the thermal bath. In this paper, we calculate
the decay rate of the heavy neutrino in a consistent way
(chapter~\ref{decay}) which automatically includes the effect of
leptonic quasiparticles, compare it to the conventional approach (chapter 
\ref{leptogenesis}) and give an outlook of what needs to be done to
arrive at consistent descriptions of leptogenesis. Our calculation is
general enough to be applied to all decays of a Yukawa
fermion at finite temperature, which has interesting implications for
other early universe dynamics (chapter \ref{conclusions}).

\section{Hard thermal loops and thermal masses}

If a particle reaction like scattering or decay takes place in the
background of a heat bath, e.g.~in the hot state of the early
Universe, thermal field theory has to be employed to describe this
process. There are two different approaches for considering finite
temperatures within quantum field theory, the imaginary and real time
formalism \cite{LeBellac:1996}, both yielding the same results. In
this work, we will use the imaginary time formalism. Going from zero
to finite temperature, ensemble-weighted expectation values of
operators have to be used rather than vacuum expectation values. For
an operator $\hat{A}$, this reads
\begin{equation}
\langle \hat{A} \rangle_\beta = {\rm tr} (\rho \hat{A}),
\end{equation}
where $\rho$ is the density operator describing the ensemble. In this
way it can be shown that the propagator at finite temperature $T$ is
given by its usual vacuum expression where the zero component of the
momentum is replaced by imaginary discrete Matsubara frequencies $q_0
= 2ni\pi T$ in the case of bosons or $(2n+1)i\pi T$ in the case of
fermions with integers $n$ (see e.g.~\cite{Thoma:2000}). Perturbation
theory at finite temperature then follows from using these propagators
and summing over the Matsubara frequencies in loop diagrams.

However, using these bare thermal propagators can lead to inconsistent
results, which are not complete to leading order, infrared divergent,
and gauge dependent in the case of gauge theories. A famous example is
the damping rate of a plasma wave in the quark gluon plasma, which is
different in different gauges. In order to cure this, the hard thermal
loop (HTL) resummation has been invented
\cite{Braaten:1989mz,Braaten:1990az}. For this purpose one has to
distinguish between hard momenta of the order $T$ or larger and soft
momenta of the order $gT$ or smaller, where $g$ is the coupling
constant, which is strictly possible only in the weak coupling limit
$g \ll 1$. After all, the HTL improved perturbation theory has been
successfully applied to thermal QCD for the description of the
quark-gluon plasma (see e.g.~\cite{Thoma:1995}).  The basic idea is
that the bare propagators are replaced by resummed propagators, if the
external momentum is soft $Q \lesssim g T$.

\begin{figure}
\begin{center}
\psfrag{d}{}
\includegraphics[scale=0.7]{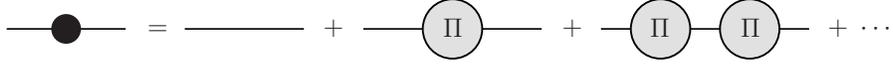}
\caption{\label{resummed}Resummed propagator}
\end{center}
\end{figure}

\pagebreak
For a scalar field (Fig.~\ref{resummed}), this resummation follows
from the Dyson-Schwinger equation as
\begin{equation}
i \Delta^*=i \Delta+i \Delta (-i \Pi) i \Delta + \dots =
\frac{i}{\Delta^{-1} -\Pi} = \frac{i}{Q^2-m_0^2-\Pi}.
\end{equation}
The thermal self-energy $\Pi$ of the scalar field then acts as a
thermal mass $m_{\rm th}^2=\Pi$ and gives a correction to the
zero-temperature mass $m_{\rm tot}^2=m_0^2+m_{\rm th}^2$. Since $\Pi$
is of the order $\sim gT$, the resummation will only affect the
propagator when $Q \lesssim \Pi \sim gT$, which is reflected in the
prescription to resum only soft momenta. The resummed fermion
propagator has a more complicated structure and will be explained in
the next section. In general, the self-energy is momentum dependent,
e.g.~the photon self-energy in QED. In this case, the leading order
gauge independent self-energy follows from integrating only over
hard momenta in the loop diagram defining the self-energy. This
HTL contribution in the resummed propagator leads to a correction
of the order $gT$ which cannot be neglected if the momentum
of the propagator is soft. The poles of the HTL resummed propagators
then describe the dispersion relations in the medium, e.g. plasma
waves following from the resummed photon propagator. In addition to 
propagators also HTL effective vertices related to the propagators
by Ward identities might have to be used. 

\section{Decay and inverse decay rate}
\label{decay}

In the neutrino decay we want to calculate, the Higgs boson and the
lepton acquire thermal masses of the order $m_{\phi,\ell}
\sim 0.2-0.4 \; T$ via their interactions with other
standard model particles. In the regime where the temperature is of
the order of the neutrino mass $T \sim M$, one of the momenta of the
decay products can be soft and has to be resummed. In the regime where
$M \lesssim 0.2-0.4 \; T$, both Higgs boson and lepton momentum will
be soft and need to be resummed. We are interested in both regimes,
therefore we will resum both Higgs boson and lepton propagator. The case of
resumming only one propagator is included in this approach, since
resumming a hard propagator gives only a negligible correction to the
bare propagator.  The HTL resummation has been invented for the weak
coupling limit $g \ll 1$. This limit does not apply in our case, our
more phenomenological approach is rather motivated by the desire to
capture effects beyond perturbation theory and justified a posteriori
by the sizeable corrections it reveals, similar to the treatment of
meson correlation functions in \cite{Karsch:2000gi}.

We consider a leptogenesis-inspired model with a massive Majorana
fermion {\sl N} coupling to a massless Dirac fermion $\ell$ and a
massless scalar $\phi$. The interaction and mass part of the
Lagrangian then reads
\begin{equation} 
\label{L}
\mathcal{L}_{\rm int,mass}=
g \bar{N} \phi \ell - \frac{1}{2} M \bar{N} N^c+h.c.\: , 
\end{equation} 
The HTL resummation technique has been considered in \cite{Thoma:1994yw}
for the case of a Dirac fermion with Yukawa coupling, from which the
HTL resummed propagators for the Lagrangian in Eq.~\eqref{L} follow directly.
We like to calculate the interaction rate $\Gamma$ of $N
\leftrightarrow \ell \phi$.

\begin{figure}
\begin{center}
\includegraphics[scale=0.5]{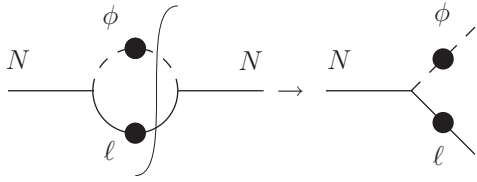}
\caption{\label{cut} $N$ decay via the optical theorem with dressed 
propagators denoted by 
a blob}
\end{center}
\end{figure}

We cut the $N$ self energy and use the 
HTL resummation for the fermion and scalar propagators (Fig.~\ref{cut}).

According to finite-temperature cutting rules
\cite{Weldon:1983jn,Kobes:1986za}, the interaction rate reads
\begin{equation}
\Gamma(P) = - \frac{1}{2 p_0} \; {\rm tr} [ (\slashed{P}+M) \; {\rm Im} \; 
\Sigma(P)].
\end{equation}
At finite temperature, the self-energy reads
\begin{equation}
\Sigma(P)=-g^2 T \sum_{k_0=i (2 n+1) \pi T} \int \frac{{\rm d}^3 k}{(2
\pi)^3} \: P_L \: S^*(K) \: P_R \: D^*(Q),
\end{equation} 
where $P_L$ and $P_R$ are the projection operators on left- and
right-handed states, $Q=P-K$ and we have summed over Majorana and Dirac spins.

The HTL-resummed scalar propagator is
\begin{equation}
D^*(Q)=\frac{1}{Q^2-m_\phi^2},
\end{equation}
where $m_\phi^2=g^2 T^2 / 12$ is the thermal mass of the scalar, created
by the interaction with fermions. Due to the reduced Majorana
degrees of freedom, $m_\phi$ differs from the Dirac-Dirac case by a
factor 1/2 \cite{Thoma:1994yw}.

The effective fermion propagator in the helicity-eigenstate representation is 
given by
\cite{Braaten:1990wp,Kapusta:1991qp,Braaten:1992gd}
\begin{equation}
\label{fermprop}
S^*(K)=\frac{1}{2} \Delta_+(K) (\gamma_0-\hat{\bf k} \cdot
\boldsymbol{\gamma}) +\frac{1}{2} \Delta_-(K) (\gamma_0+\hat{\bf k} \cdot
\boldsymbol{\gamma}),
\end{equation}
where 
\begin{equation}
\Delta_\pm(K)=\left [ -k_0 \pm  k + \frac{m_\ell^2}{k} \left ( \pm1 - 
\frac{\pm k_0 - k}{2k} \ln \frac{k_0+k}{k_0-k}  \right ) \right ]^{-1}
\end{equation}
and
\begin{equation}
m_\ell^2=\frac{1}{32} g^2 T^2.
\end{equation}
This again differs from the Dirac case by a factor 1/2 \cite{Thoma:1994yw}.

The trace can be evaluated as
\begin{equation}
{\rm tr} [(\slashed{P} +M) P_L S^*(K) P_R]= \Delta_+ (p_0-p \eta)+\Delta_-
(p_0 +p \eta),
\end{equation}
where $\eta=\cos\theta$ is the angle between {\bf p} and {\bf k}. We
evaluate the sum over Matsubara frequencies by using the Saclay method
\cite{Pisarski:1987wc}. For the scalar propagator, the Saclay
representation reads
\begin{equation}
D^*(Q)=-\int_0^\beta {\rm d} \tau e^{q_0 \tau} \frac{1}{2 \omega_q}
\{ [ 1+n_B(\omega_q)] e^{-\omega_q \tau} + n_B(\omega_q) e^{\omega_q \tau}\},
\end{equation}
where $\beta=1/T$, $n_B(\omega_q)=1/(e^{\omega_q \beta}-1)$
is the Bose-Einstein distribution and $\omega_q^2=q^2+m_\phi^2$. For
the fermion propagator it is convenient to use the spectral
representation \cite{Pisarski:1989cs} 
\begin{equation}
\Delta_\pm(K)=-\int_0^\beta {\rm d \tau'} e^{k_0 \tau'} 
\int_{-\infty}^{\infty}
{\rm d} \omega \: \rho_\pm(\omega,k) [1-n_F(\omega)] e^{-\omega
\tau'},
\end{equation} 
where $n_F(\omega)=1/(e^{\omega \beta}+1)$ is the Fermi-Dirac
distribution and $\rho_\pm$ the spectral density
\cite{Braaten:1990wp,Kapusta:1991qp}.

The fermion propagator
in Eq.~\eqref{fermprop} has two different poles for $1/\Delta_\pm=0$, which
correspond to two leptonic quasiparticles with a positive
($\Delta_+$) or negative ($\Delta_-$) ratio of helicity over chirality
\cite{Klimov:1981ka,Weldon:1982aq,Weldon:1982bn,Weldon:1989ys}. The
spectral density $\rho_\pm$ has two contributions, one from the poles
and one discontinuous part. Since the quasi-particles are our final
states, we will set $K$ such that $1/\Delta_\pm(K)=0$. 

\begin{figure}
\begin{center}
\includegraphics[scale=0.93]{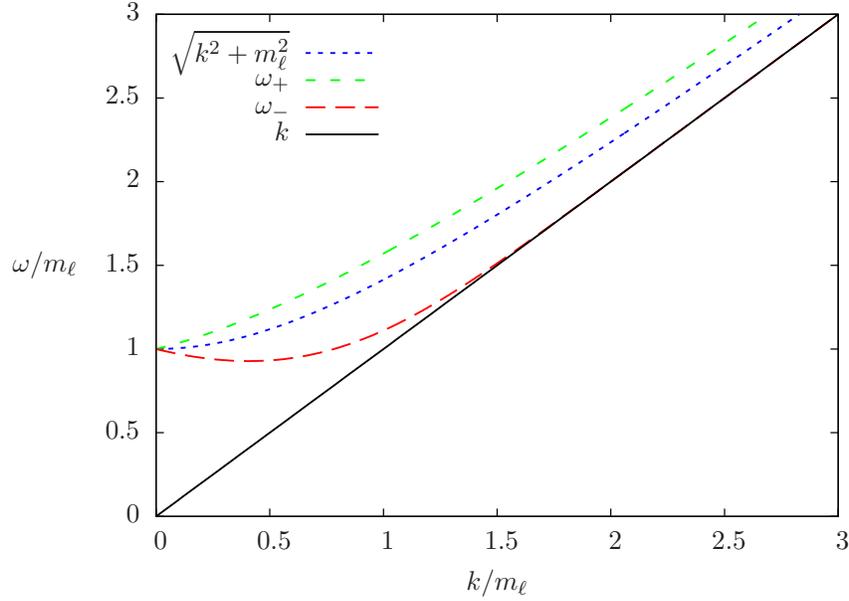}
\caption{\label{modes} The two leptonic dispersion relations compared with the 
standard
dispersion relation $\omega^2=k^2+m_\ell^2$ in blue are shown.}
\end{center}
\end{figure}

\begin{figure}
\begin{center}
\includegraphics[scale=0.95]{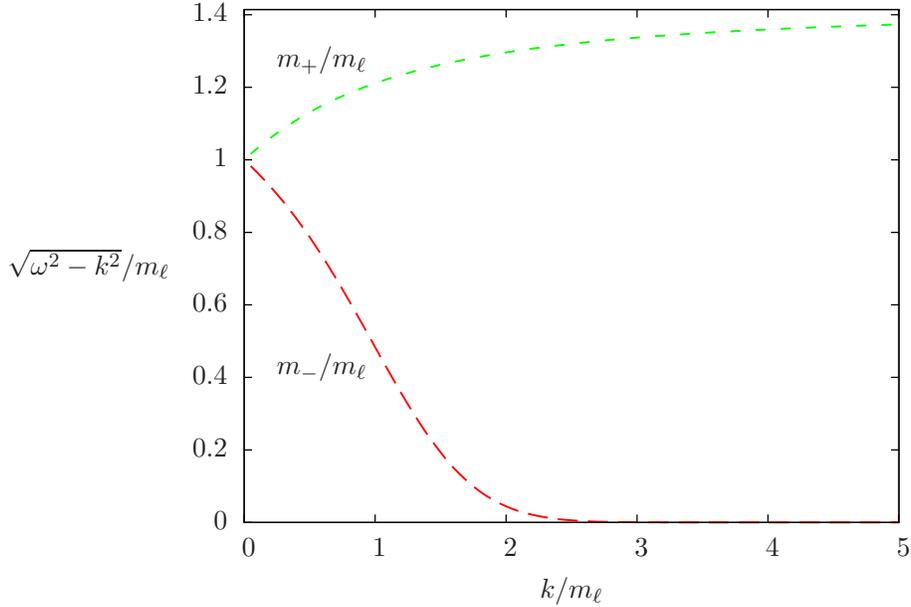}
\caption{\label{masses} The momentum-dependent quasiparticle masses 
$m_\pm^2=\omega_\pm^2-k^2$ are shown.}
\end{center}
\end{figure}

Thus we are
only interested in the pole contribution
\begin{equation}
\label{rho}
\rho_\pm^{\rm pole}(\omega,k)=\frac{\omega^2-k^2}{2 m_\ell^2} (\delta(\omega-
\omega_\pm)+ \delta (\omega+ \omega_\mp)),
\end{equation}
where $\omega_\pm$ are the dispersion relations for the two
quasiparticles, i.e.~the solutions for $k_0$ such that
$1/\Delta\pm(\omega_\pm,{\bf k})=0$, shown in Fig.~\ref{modes}. There
exists an analytical solution for $\omega_\pm$ making use of the Lambert
W function which has not yet been reported in the literature. The
analytical solution is explained in detail in the appendix. One can
assign a momentum-dependent thermal mass
\mbox{$m_\pm(k)^2=\omega_\pm(k)^2-k^2$} to the two modes as shown in
Fig.~\ref{masses} and for very large momenta the heavy mode $m_+$
approaches $\sqrt{2} \: m_\ell$, while the light mode becomes
massless.

In order to execute the sum over Matsubara frequencies, we write
$k_0=i \omega_n$ with $\omega_n=(2 n+1) \pi T$ and remember that when
evaluating frequency sums, also $p_0=i \omega_m=i (2 m + 1) \pi T$ can
be written as a Matsubara frequency and later on be continued
analytically to real values of $p_0$
\cite{LeBellac:1996,Baym,Dolan:1973qd}. In particular $e^{p_0
\beta}=e^{i \omega_m \beta}=-1$. We can write
\begin{equation}
T \sum_n e^{i \omega_n \tau}= \sum_{n'=-\infty}^{\infty}
\delta(\tau-n' \beta), 
\end{equation} 
then
\begin{equation}
T \sum_n e^{(p_0-k_0) \tau} e^{k_0 \tau'}=e^{p_0 \tau} \delta(\tau'-\tau),
\end{equation}
since $-\beta \leq \tau'-\tau \leq \beta$.  After evaluating the sum
over $k_0$ and carrying out the integrations over $\tau$ and $\tau'$,
we get
\begin{equation}
\begin{split}
T \sum_{k_0} D^*(Q) \Delta_\pm(K)=
 -\int_{-\infty}^{\infty} {\rm d} \omega \,
\rho_\pm(\omega,k) \frac{1}{2 \omega_q} & \left [ 
\frac{1+n_B(\omega_q)-n_F(\omega)}{p_0 - \omega -\omega_q} \right. \\
& \left. \quad +\frac{n_B(\omega_q)+n_F(\omega)}{p_0 - \omega +\omega_q} \right ].
\end{split}
\end{equation}
Integrating $\omega$ over the pole part of $\rho_\pm$ in
Eq.~\eqref{rho}, we get
\begin{equation}
\begin{split}
T \sum_{k_0} D^* \Delta_\pm=-\frac{1}{2 \omega_q} 
& \left \{ 
\frac{\omega_\pm^2-k^2}
{2 m_\ell^2} \left [ 
\frac{1+n_B-n_F}{p_0-\omega_\pm-\omega_q}+
 \frac{n_B+n_F}{p_0-\omega_\pm+\omega_q} 
\right ]
\right. \\
& 
\label{frequencysum}
\left. +  \frac{\omega_\mp^2-k^2}{2 m_\ell^2} \left [ 
\frac{n_B+n_F}{p_0+\omega_\mp-\omega_q}+
 \frac{1+n_B-n_F}{p_0+\omega_\mp+\omega_q} \right ] 
\right \}
\end{split}
\end{equation} 
where $n_B=n_B(\omega_q)$ and $n_F=n_F(\omega_\pm)$ or
$n_F(\omega_\mp)$, respectively.

The four terms in Eq.~\eqref{frequencysum} correspond to the processes
with the energy relations indicated in the denominator, i.e.~the decay
$N \rightarrow \phi \ell$, the production $N \phi \rightarrow \ell$,
the production $N \ell \rightarrow \phi$ and the production of $N \ell
\phi$ from the vacuum, as well as the four inverse reactions
\cite{Weldon:1983jn}. We are only interested in the process $N \leftrightarrow 
\phi \ell$, where the decay and inverse decay are illustrated by the
statistical factors
\begin{equation}
1+n_B-n_F=(1+n_B)(1-n_F)+n_B n_F.
\end{equation}
Our term thus reads
\begin{equation}
T \sum_{k_0} D^* \Delta_\pm = - \frac{1}{2 \omega_q} \;
\frac{\omega_\pm^2-k^2}{2 m_\ell^2} \; 
\frac{1+n_B-n_F}{p_0-\omega_\pm-\omega_q}.
\end{equation}

For carrying out the integration over the angle $\eta$, we use
\begin{equation}
{\rm Im} \frac{1}{p_0-\omega_\pm-\omega_q}= - \pi
\delta(p_0-\omega_\pm-\omega_q) = - \pi \frac{\omega_q}{k p}
\delta(\eta-\eta_\pm), 
\end{equation}
where
\begin{equation}
\eta_\pm=\frac{1}{2 k p} \left [ 2 p_0 \omega_\pm - M^2 - (\omega_\pm^2-k^2)
+ m_\phi^2 \right ]
\end{equation}
denotes the angle for which the energy conservation
$p_0=\omega+\omega_q$ holds. The integration over $\eta$ then yields
\begin{equation} 
\int_{-1}^{1} {\rm d} \eta \; {\rm Im}(T \sum_{k_0} D^*
\Delta_\pm)= \frac{\pi}{2 k p} \; \frac{\omega_\pm^2-k^2}{2 m_\ell^2} \; 
[1+n_B(\omega_{q \pm})-n_F(\omega_\pm)],
\end{equation}
where $\omega_{q \pm}=p_0-\omega_\pm$.
It follows that
\begin{equation}
\begin{split}
\Gamma(P)=&-\frac{1}{2 p_0}\; {\rm tr} [(\slashed{P} +M) {\rm Im} \; \Sigma(P)] 
\\
=& \frac{1}{2 p_0} \; {\rm Im} \left \{ g^2 T \sum_{k_0} \int
\frac{{\rm d^3} k}{(2 \pi)^3} \; {\rm tr} [(\slashed{P} +M) P_L S^* P_R] D^*
\right \} \\
=& \frac{g^2}{8 \pi^2 p_0} \; {\rm Im} \left \{ T \sum_{k_0} \int {\rm d} k \,
{\rm d} \eta \: k^2 D^* [\Delta_+ (p_0-p \eta) +\Delta_-
(p_0 +p \eta)] \right \} \\
=& \frac{g^2}{32 \pi p_0 p} \sum_\pm \int_{-1 \leq \eta_\pm \leq 1} {\rm d} k 
\: \frac{\omega_\pm^2-k^2}{2 m_\ell^2}
[1+n_B(\omega_{q \pm}) -n_F(\omega_\pm)] \\ 
&  \hspace{3.4cm} \times [2 p_0
(k \mp \omega_\pm) \pm M^2 \pm (\omega_\pm^2-k^2) \mp m_\phi^2],
\end{split}
\end{equation}
where we only integrate over regions with $-1 \leq \eta \leq 1$.

Using finite temperature cutting rules, one can also write the
interaction rates for the two modes in a way which resembles the
zero-temperature case \cite{Weldon:1983jn}
\begin{equation}
\begin{split}
\Gamma_\pm(P)=\frac{1}{2 p_0}\int \, {\rm d} \tilde{k} \, {\rm d} \tilde{q} \; 
& (2 \pi)^4 \delta^4(P-K-Q) \; |\mathcal{M}_\pm(P,K)|^2 \\
 \times &  [1+n_B(\omega_q)-n_F(\omega_\pm)],
\end{split}
\end{equation}
where 
\begin{equation}
{\rm d} \tilde{k}=\frac{{\rm d}^3 k}{(2 \pi)^3 2 \,k_0}
\end{equation}
and ${\rm d}\tilde{q}$ analogously and the matrix elements are
\begin{equation}
|\mathcal{M}_\pm(P,K)|^2=g^2 \frac{\omega_\pm^2-k^2}{2 m_\ell^2} \omega_\pm 
\left (p_0 \mp p \eta_\pm \right ). 
\end{equation}

Now that we have arrived at an expression for the full HTL decay rate
of a Yukawa fermion, we would like to compare it to the
conventional approximation adopted by \cite{Giudice:2003jh}. To this
end, we do the same calculation for an approximated
fermion propagator
\begin{equation}
\label{approx}
S^*_{\rm 
approx}(K)=\frac{1}{\slashed{K}-m_\ell}
\end{equation}
This yields the following approximated interaction rate:
\begin{equation}
\begin{split}
\Gamma_{\rm approx}(P)=& \frac{g^2}{32 \pi p_0 p} \int_{k_1}^{k_2} {\rm d} k 
\: 
\frac{k}{\omega} [1+n_B(\omega_q) -n_F(\omega)] [M^2 +m_\ell^2 -m_\phi^2] 
\\
=& \frac{1}{2 p_0}\int \, {\rm d} \tilde{k} \: {\rm d} \tilde{q} \: 
  (2
\pi)^4 \delta^4(P-K-Q) |\mathcal{M}|^2 \\ 
& \hspace{1.6cm} \times 
[1+n_B(\omega_q)-n_F(\omega)],
\end{split}
\end{equation}
where $\omega^2=k^2+m_\ell^2$, $\omega_q=p_0-\omega$ and the
integration boundaries
\begin{equation}
k_{1,2}= \frac{1}{2 M^2} \left [p_0 \sqrt{(M^2+m_\ell^2-m_\phi^2)^2-
(2 M m_\ell)^2} \mp p (M^2+m_\ell^2-m_\phi^2) \right ]
\end{equation}
ensure $-1 \leq \eta \leq 1$, where 
\begin{equation}
\eta=\frac{1}{2 k p} \left [ 2 p_0 \omega - M^2 - m_\ell^2
+ m_\phi^2 \right ].
\end{equation}
We see that the matrix element is
\begin{equation}
|\mathcal{M}|^2=\frac{g^2}{2} (M^2+m_\ell^2-m_\phi^2).
\end{equation}

This result resembles the zero temperature result
\begin{equation}
\Gamma_{T=0}(P)= \frac{g^2}{32 \pi p_0 p} \int_{k_1}^{k_2} {\rm d} k \: 
 \frac{k}{\omega} 
\left [ M^2 +m_\ell^2 -m_\phi^2 \right ]
\end{equation}
with zero temperature masses $m_\ell$, $m_\phi$.
The missing factor
\begin{equation}
1+n_B-n_F=(1+n_B)(1-n_F)+n_B n_F
\end{equation}
accounts for the statistical distribution of the initial or final
particles.  As pointed out in more detail in \cite{Kiessig:2009cm}, we
have shown that the approach to treat thermal masses like zero
temperature masses in the final state \cite{Giudice:2003jh} is
justified since it equals the HTL treatment with an approximate
fermion propagator. However this approach does not equal the full HTL
result.

Concluding this calculation, a caveat has to be added: The external
Majorana fermion will also acquire a thermal mass of order
$gT$. Thus, if its zero temperature mass is smaller than that, the
external fermion also needs to be described by leptonic quasiparticles
to be consistent. However, in our leptogenesis application, the Yukawa
coupling giving rise to the Majorana neutrino decay is much smaller
than the couplings giving rise to the thermal masses of the Higgs
boson (scalar) and the lepton (Dirac fermion) and thus the thermal
mass of the heavy neutrino can be neglected.

We have calculated the decay rate assuming a Majorana particle, but
the result can be very easily generalized to the case of two Dirac
fermions by inserting the appropriate factors of two in the decay rate
and the thermal masses.

\section{Neutrino decay in Leptogenesis}
\label{leptogenesis}

When turning to leptogenesis with
\begin{equation}
\delta {\mathcal L} = i \bar{N}_i \partial_\mu \gamma^\mu N_i - \lambda_{\nu,i
  \alpha} \bar{N}_i \phi^\dagger \ell_\alpha - \frac{1}{2} M_i
\bar{N}_i N_i^c + h.c.,
\end{equation}
we sum over the two components of the doublets, particles and
antiparticles and the three lepton flavors. Thus we need to replace
$g^2 \rightarrow 4 (\lambda_\nu^\dagger
\lambda_\nu)_{11}$. Integrating over all neutrino momenta, the decay
density in equilibrium is
\begin{equation}
\gamma_D^{\rm eq}= \int \frac{{\rm d^3} p}{(2 \pi)^3} f_N^{\rm eq}(E) \, 
\Gamma_D= \frac{1}{2 \pi^2} \int_M^\infty {\rm d} E \: E \, p \, f_N^{\rm eq} 
\, 
\Gamma_D,
\end{equation} 
where $E=p_0$, $f_N^{\rm eq}(E)=[\exp(E \beta)-1]^{-1}$ is the
equilibrium distribution of the neutrinos and $\Gamma_D=[1-f_N^{\rm
eq}(E)] \, \Gamma$.  

Since $\lambda_{\nu,i \alpha} \ll 1$, the thermal
masses are 
\begin{equation}
{m_\phi^2(T)=\left (\frac{3}{16} g_2^2+ \frac{1}{16} g_Y^2 +
\frac{1}{4} y_t^2 + \frac{1}{2} \lambda \right) T^2}
\end{equation}
and
\begin{equation}
m_\ell^2(T)=\left (\frac{3}{32} g_2^2+ \frac{1}{32} g_Y^2 \right ) T^2.
\end{equation}
The couplings denote the SU(2) coupling $g_2$, the U(1) coupling
$g_Y$, the top Yukawa coupling $y_t$ and the Higgs self coupling
$\lambda$, where we assume a Higgs mass of $115$ GeV. The other Yukawa
couplings can be neglected since they are much smaller than unity and
the remaining couplings are renormalized at the first Matsubara mode $2
\pi T$ as explained in \cite{Giudice:2003jh}.

\begin{figure}
\begin{center}
\includegraphics[scale=0.90]{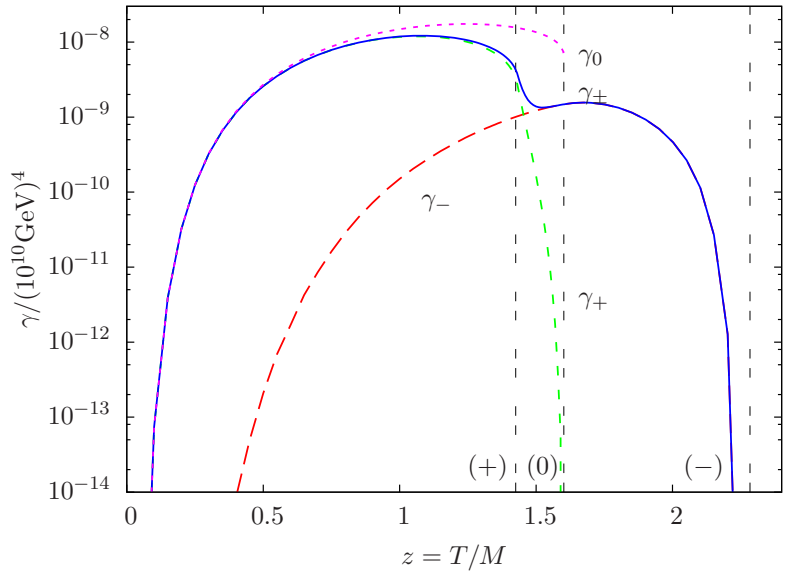}
\caption{\label{comp} The neutrino decay density with the one lepton mode 
approach $\gamma_0$ and the two-mode treatment $\gamma_\pm$ for
$M_{1}=10^{10}$ GeV and $\tilde{m}_1 = 0.06$ eV. The thresholds for
the two modes (+), (-) and one mode (0) are indicated.}
\end{center}
\end{figure}

In Fig.~\ref{comp}, we compare our consistent HTL calculation to the
approximation adopted by \cite{Giudice:2003jh}, while we add quantum
statistical distribution functions to their calculation which then
equals the approach of using an approximated lepton propagator
$1/(\slashed{K}-m_\ell)$ as in Eq.~\eqref{approx} \cite{Kiessig:2009cm}. We
assume the heavy neutrino masses to be hierarchical and evaluate the
decay rate for the typical value $M_1=10^{10}$ GeV, which is inspired by
putting $M_3$ to the GUT scale ($10^{15}$ GeV) and assuming $M_1/M_3
\sim 10^{-5}$ analogous to the quark sector. The combination of Yukawa
couplings $(\lambda_\nu \lambda^\dagger_\nu)_{11}$ which governs the
decay rate is often parametrized by the so called 'effective' neutrino
mass $\tilde{m_1}=(\lambda_\nu \lambda^\dagger_\nu)_{11} \, v^2/M_1$,
where \mbox{$v=174$ GeV} is the vacuum expectation value of the Higgs
field. We take \mbox{$\tilde{m_1}=0.06$ eV}, inspired by the mass scale of
the atmospheric mass splitting. However, our results can be
generalized to all regions of parameter space. 

In the one-mode approach, the decay is forbidden when the thermal
masses of Higgs boson and lepton become larger than the neutrino mass
$M<m_\ell+m_\phi$. Considering two modes, the kinematics exhibit a
more interesting behavior. For the positive mode, the phase space is
reduced due to the larger quasi-mass and at $M=m_+(\infty)+m_\phi$,
the decay is only possible into leptons with small momenta, thus the
rate drops dramatically. The decay into the negative, quasi-massless
mode is suppressed since its residue is much smaller than the one of the
positive mode. However, the decay is possible up to $M=m_\phi$. Due to
the various effects, the two mode rate differs from the one mode
approach by more than one order of magnitude in the interesting
temperature regime of $z=T/M \gtrsim 1$.

It is extremely tempting to put this result in a Boltzmann-solver and
obtain an effect for the produced baryon asymmetry. However, in the
quest for consistent treatments which capture effects of the same
origin and size, other effects need to be included as well. At higher
temperatures, when $m_\phi > M+m_\pm(k)$, the Higgs can decay into
neutrino and lepton modes and this process acts as production mechanism for
neutrinos \cite{Giudice:2003jh}. Moreover, the CP asymmetry needs to
be calculated taking into account the two lepton modes in order to
have a consistent treatment.

\section{Conclusions}
\label{conclusions}

As discussed in detail in \cite{Kiessig:2009cm}, we have, by employing
HTL resummation and finite temperature cutting rules, confirmed that
treating thermal masses as kinematic masses as in
\cite{Giudice:2003jh} is a reasonable approximation. However, quantum
statistical functions need to be included as they always appear in
thermal field theory. Moreover, the full HTL lepton propagator shows a
non-trivial two-mode behavior which is not accounted for by the
conventional approach. We have calculated the effect of the two modes
in a general way which is applicable to any decay and inverse decay
rates involving fermions at high temperature. Thus, this calculation
is a valuable tool for other particle processes in the early
universe, as other leptogenesis processes, the thermal production
of gravitinos or the like.  

The behavior of the decay density of the
two lepton modes can be explained by considering the dispersion
relations $\omega_\pm$ of the modes and assigning momentum-dependent
quasi-masses to them. The thresholds for neutrino decay reported in
\cite{Giudice:2003jh} are shifted and the decay density shows
deviations of more than an order of magnitude in the interesting
temperature regime $T/M \sim 1$. Thus we expect these effects to have
a sizeable impact on the final baryon asymmetry. However, in order to arrive
at a minimal consistent treatment, also the decay $\phi \rightarrow N \ell$
at high temperatures needs to be included as well as a CP asymmetry
that is corrected for lepton modes. In a further step, it will be
interesting to include the effect of thermal widths in the
calculations.

As for all effects arising from thermal field theory, the effects are
only important in the weak washout regime, where leptogenesis takes
place at high temperatures. We are aware of the progress that is
currently being made in approaching the effects of quantum
statistics
\cite{Basboll:2006yx,Garayoa:2009my,HahnWoernle:2009qn,HahnWoernle:2009en},
quantum transport equations
\cite{Anisimov:2008dz,Garny:2009rv,Garny:2009qn,Anisimov:2010aq,Garny:2010nj,Beneke:2010wd}
or other collective phenomena as e.g.~the
Landau-Pomeranchuk effect
\cite{Besak:2010fb}. These efforts contribute to getting an idea of
the size and impact of various thermal effects by approaching the
extremely complex situation from different angles.
\\[1ex]
{\bf Acknowledgements} We would like to thank Georg Raffelt, Florian
Hahn-W\"ornle, Steve Blanchet, Matthias Garny, Marco Drewes, Wilfried
Buchm\"uller and Annika W\"ohner for fruitful and inspiring discussions.

\appendix

\section{Analytical solution for HTL lepton dispersion relations}

The dispersion relations of the two lepton modes are given by the
poles of the corresponding propagator. Hence, we seek the zeros of
\begin{equation}
\label{D}
D_\pm(K)=\Delta_\pm(K)^{-1}=\left [ -k_0 \pm k + \frac{m_\ell^2}{k}
  \left ( \pm1 - \frac{\pm k_0 - k}{2k} \ln \frac{k_0+k}{k_0-k} \right
  ) \right ]^{-1}
\end{equation}
The equations $D_\pm=0$ can be transformed by the substitutions
\begin{align}
x_+  &:=  \frac{k_0+k}{k_0-k} \\
x_- &:= \frac{k_0-k}{k_0+k}=\frac{1}{x_+} \\
c &:= \frac{k^2}{m_\ell^2}.
\end{align}
This yields
\begin{equation}
D_\pm=\pm \frac{k}{c} \frac{1}{x_\pm-1} \left (-2 c-1+x_\pm-\ln x_\pm \right).
\end{equation}
Further introducing
\begin{equation}
\label{s}
s := - \exp (-2 c -1 )
\end{equation}
leads to
\begin{equation}
D_\pm= \frac{\mp 2 k}{1+\ln(-s)} \frac{1}{x_\pm-1} \left [ x_\pm +
  \ln(-s) - \ln x_\pm \right ].
\end{equation}
Since the prefactor does not have poles for the values of $K$ we are
looking at, solving $D_\pm=0$ amounts to solving 
\begin{equation} x_\pm
+\ln(-s)-\ln x_\pm=0, 
\end{equation} 
which in turn means 
\begin{equation} s=-x_\pm
\e^{-x_\pm}.  
\end{equation}

This is the defining equation of the Lambert W function
\cite{Corless:1996zz,Chapeau}, thus the solution reads
\begin{equation}
x_\pm=-W(s).
\end{equation}
According to the definition in Eq.~\eqref{s}
\begin{equation}
\label{srange}
-1/\e \leq s \leq 0,
\end{equation}
thus the two real branches of the Lambert function, $W_0$ and
$W_{-1}$, correspond to the two solutions we seek. In the range given
by Eq.~\eqref{srange} $W_0 \geq -1$ and $W_{-1} \leq -1$. For $k_0 \geq
k$ we have $x_+ \geq 1$ and $x_- \leq 1$. Hence, the physical solutions for $x_\pm$ read
\begin{equation}
x_+=-W_{-1}(s) \hspace{1cm} {\rm and} \hspace{1cm} x_-=-W_0(s).
\end{equation}

The corresponding results for $\omega_\pm$ are then given by
\begin{align}
\omega_+&= k \; \frac{W_{-1}(s) -1}{W_{-1}(s)+1} \label{op} \\
\omega_-&=- k \; \frac{W_{0}(s) -1}{W_{0}(s)+1}. \label{om}
\end{align}
Making use of the relations \cite{Jeffrey}
\begin{equation}
W_{0,-1}(z)+\ln (W_{0,-1}(z))=\ln z,
\end{equation}
one can directly prove the result by plugging Eqs.~\eqref{op} and
\eqref{om} into Eq.~\eqref{D}.

\end{document}